\documentclass[aps,amssymb,10pt,showpacs,showkeys,letterpaper]{revtex4}
\usepackage[utf8]{inputenc}
\usepackage{graphicx}
\usepackage{amsmath}
\usepackage{epsfig}
\usepackage{bm}

\newcommand{\be}{\begin{equation}}
\newcommand{\ee}{\end{equation}}
\newcommand{\beq}{\begin{eqnarray}}
\newcommand{\eeq}{\end{eqnarray}}

\begin{document}

\title{
Motion and trajectories of photons in  a three-dimensional\\ rotating Ho\v{r}ava AdS black hole
 }
\author{P. A. Gonz\'{a}lez}
\email{pablo.gonzalez@udp.cl}
\affiliation{Facultad de
	Ingenier\'{i}a y Ciencias, Universidad Diego Portales, Avenida Ej\'{e}rcito
	Libertador 441, Casilla 298-V, Santiago, Chile.}
\author{ Marco Olivares }
\email{marco.olivaresr@mail.udp.cl}
\affiliation{ Facultad de Ingenier\'ia y Ciencias, Universidad Diego Portales,
	Avenida Ej\'ercito Libertador 441, Casilla 298-V, Santiago, Chile.}
\author{Eleftherios Papantonopoulos}
\email{lpapa@central.ntua.gr}
\affiliation{Department of
	Physics, National Technical University of Athens,\\ Zografou Campus
	GR 157 73, Athens, Greece.}
\author{Yerko V\'{a}squez}
\email{yvasquez@userena.cl}
\affiliation{Departamento de F\'{\i}sica, Facultad de Ciencias, Universidad de La Serena,\\
	Avenida Cisternas 1200, La Serena, Chile.}

\date{\today}

\begin{abstract}

We study the motion of photons in the background of a three-dimensional rotating Ho\v{r}ava AdS black hole that corresponds to a Lorentz-violating version of the BTZ black hole and we analyze the effect of the breaking of Lorentz invariance on the null geodesics structure, by solving analytically the equations of motion. Mainly we find that, through a fine tuning  of the parameters of the theory, new kinds of orbits are allowed, such as unstable circular orbits and trajectories of first kind. Also, we show that an external observer will see that photons arrive at spatial infinity in a finite coordinate time.

\end{abstract}

\maketitle

\newpage

\tableofcontents

\section{Introduction}

The three-dimensional models of gravity have been of great interest, because they are much simpler than the four-dimensional and higher-dimensional models of gravity and one can investigate more efficiently some of their properties which are shared by their higher dimensional analogs, and also exhibit very interesting solutions such as particle-like solutions and black holes. The well known Bañados–Teitelboim–Zanelli (BTZ) black hole \cite{Banados:1992wn} is a solution to the Einstein equations in three dimensions with a negative cosmological constant, which presents interesting properties at both classical and quantum levels and it shares several features of the Kerr black hole \cite{Carlip:1995qv}. However, three-dimensional general relativity (GR) has no local gravitational degrees of freedom. Another model known as topologically massive gravity (TMG) was constructed by adding a Chern-Simons gravitational term to the action of three-dimensional GR \cite{Deser:1981wh}. In contrast to GR in three dimensions this model contains a propagating degree of freedom which corresponds to a massive graviton \cite{Deser:1981wh, Deser:1982vy}, and also admits the BTZ (and other) black holes as exact solutions.

However, the interest on TMG relies on the possibility of constructing a chiral theory of gravity at a special point of the space of parameters \cite{Deser:1982vy}. Some interesting solutions of the TMG theory has been found in Refs. \cite{Moussa:2003fc, Garbarz:2008qn, Vasquez:2009mk}. On the other hand, Bergshoeff, Hohm and Townsend introduced another three-dimensional massive gravity theory, which is known as BHT massive gravity, where the action is the standard Einstein-Hilbert term with a specific combination of scalar curvature square term and Ricci tensor square one \cite{Bergshoeff:2009hq, Bergshoeff:2009aq, Bergshoeff:2009tb, Bergshoeff:2009fj,Andringa:2009yc, Bergshoeff:2010mf}, and it is equivalent at the linearized level to the (unitary) Fierz-Pauli action for a massive spin-2 field \cite{Bergshoeff:2009hq}. The model in three dimensions is indeed unitary in the tree-level, but the corresponding model in higher dimensions is not so due
to the appearance of non-unitary massless spin-2 modes \cite{Nakasone:2009bn}. Also, BHT gravity admits exact solutions such as warped AdS black holes \cite{Clement:2009gq}, AdS waves \cite{AyonBeato:2009yq, Clement:2009ka}, asymptotically Lifshitz black holes
\cite{AyonBeato:2009nh}, gravitational solitons, kinks and wormholes \cite{Oliva:2009ip}. For further aspects of BHT gravity see \cite{Kim:2009jm,Oda:2009ys, Liu:2009pha, Nakasone:2009vt, Deser:2009hb}.

Nowadays, one could think that Lorentz invariance may not be fundamental or exact, but is merely an emergent symmetry on sufficiently large distances or low energies. It has been suggested in Ref. \cite{Horava:2009uw} that giving up Lorentz invariance by introducing a preferred foliation and terms that contain higher-order spatial derivatives can lead to significantly improved UV behavior, the corresponding gravity theory is dubbed Hořava gravity. The three-dimensional Hořava gravity \cite{Sotiriou:2011dr} admits a Lorentz-violating version of the BTZ black hole, i.e. a black hole solution with AdS asymptotics, only in the sector of the theory in which the scalar degree of freedom propagates infinitely fast \cite{Sotiriou:2014gna}. Remarkably, in contrast to GR, the three-dimensional Hořava gravity also admits black holes with positive and vanishing cosmological constant.
It was shown that the propagation of a massive scalar field is stable in the background of a rotating three-dimensional Ho\v{r}ava AdS black hole and by employing the holographic principle it was found that for different relaxation times the  perturbed system  reach thermal equilibrium  in the various branches of solutions \cite{Becar:2019hwk}.

An important issue in gravitational physics is to know the kind of orbits that test particles follow outside the event horizon of a black hole. This information can be provided by studying the geodesics around these black holes. In this work we will consider as background a three-dimensional rotating Ho\v{r}ava AdS black hole \cite{Sotiriou:2014gna}. Our main motivation is, by calculating the null geodesic structure, to analyze the effect of the Lorentz breaking symmetry on the motion of particles. For the BTZ background it was shown that while massive particles always fall into the event horizon and no stable orbits are possible \cite{Farina:1993xw}, massless particles can escape or plunge to the horizon \cite{Cruz:1994ir}. Also, for the three-dimensional BHT gravity, the null geodesic structure in a static circularly symmetric black hole was studied and it was shown that the deflection angle can be positive, negative or even zero, where the negative deflection angle indicates the repulsive behavior of the gravity which comes from the gravitational hair parameter that is the most characteristic quantity of the BHT massive gravity \cite{Nakashi:2019jjj}. For Lifshitz black holes in (2+1)-dimensions, the null geodesic structure does not allow the bending of the light \cite{Cruz:2013ufa}.

Additionally, the behaviour of null geodesics has been used to calculate the absorption cross section for massless scalar waves at high frequency limit or geometric optic limit, because at high frequency limit the absorption cross section can be approximated by the geometrical cross section of the black hole photon sphere $\sigma \approx \sigma_{geo}=\pi b_c^2$, where $b_{c}$ is the impact parameter of the unstable circular orbit of photons. Moreover, in \cite{Decanini:2011xi, Decanini:2011xw} this approximation was improved at the high frequency limit by $\sigma \approx \sigma_{geo}+\sigma_{osc}$, where $\sigma_{osc}$ is a correction involving the geometric characteristics of the null unstable geodesics lying on the photon sphere, such as the orbital period and Lyapunov exponent. This approximation was used recently in \cite{Fernando:2017qrd} to evaluate the absorption cross section of electromagnetic waves at high frequency limit.

The work is organized as follows. In Section \ref{Background} we give a brief review of three-dimensional rotating Ho\v{r}ava AdS black hole. In Section \ref{ME} we find the motion equations for particles and we present the null geodesic structure in Section \ref{geodesic}. Finally, our conclusions are in Section \ref{conclusion}.

\section{Three-dimensional rotating Ho\v{r}ava AdS black holes}
\label{Background}
\begin{figure}[!h]
\begin{center}
\includegraphics[width=60mm]{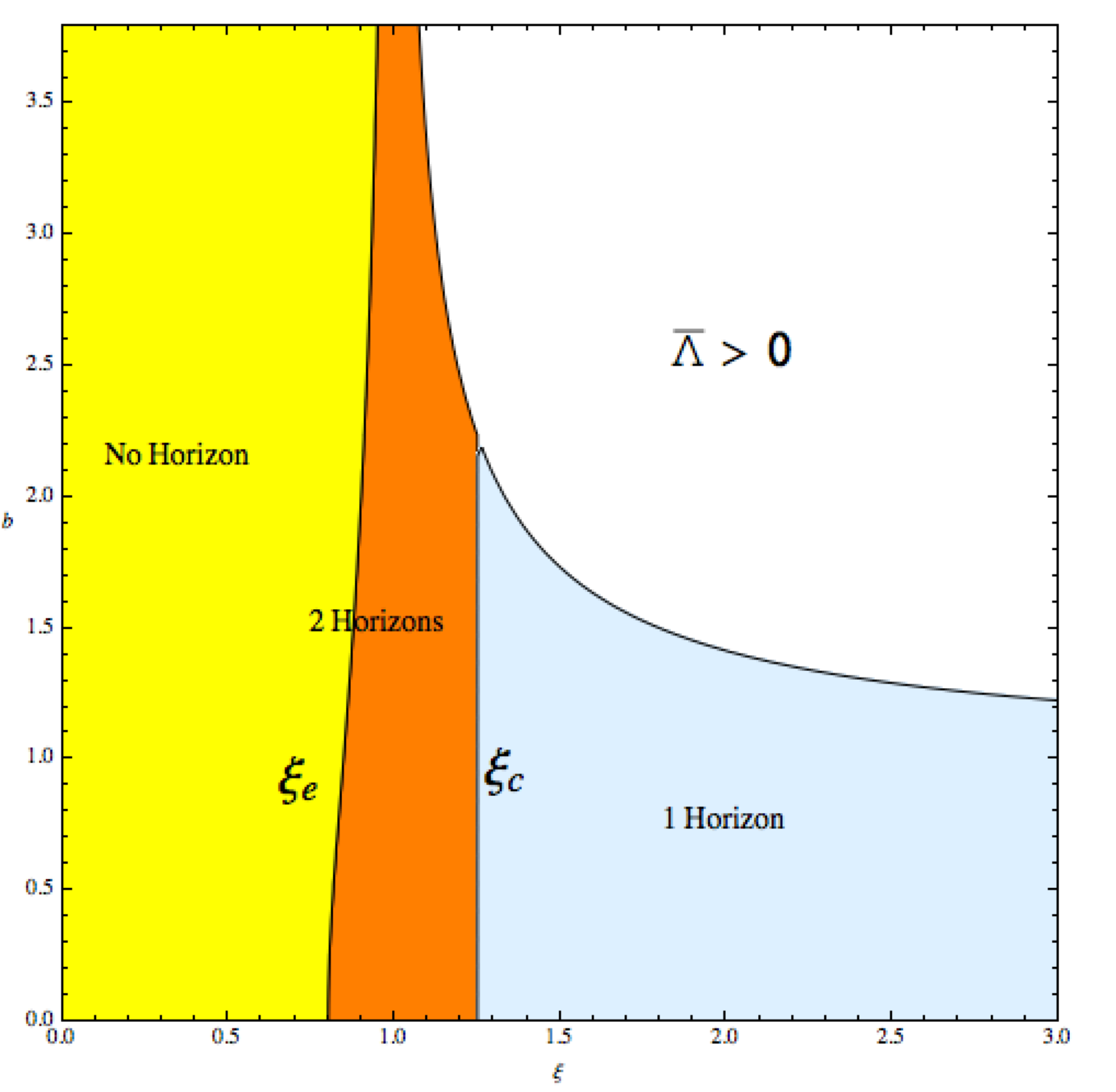}
\end{center}
\caption{Different regions of the parameter space for $a=1$, $\lambda=1$, $\Lambda=-1$, $M=1.5$ and $J=1$. The colored region corresponds to $\bar{\Lambda} <0$. In the yellow region, there are no black hole solutions. In the orange region, the black holes have two horizons while in the light blue region, they have only one horizon \cite{Becar:2019hwk}. }
\label{co}
\end{figure}

 %\textbf{I think we need to write down the basic action of the Horava gravity:\\
 %\\
The three-dimensional Ho\v{r}ava gravity is described by the action  \cite{Sotiriou:2011dr}
\begin{equation}
S_{H}=\frac{1}{16\pi G_{H}}\int dT d^2x N\sqrt{g}\left[L_{2}+L_{4}\right] \,,
\end{equation}
where $G_{H}$ is a coupling constant with dimensions of a length squared and the Lagrangian $L_{2}$ has the following form
\begin{equation}
 L_{2}=K_{ij} K^{ij}-\lambda K^2+\xi\left(^{(2)}R-2\Lambda\right)+\eta a_{i} a^{i}\,,
\end{equation}
where $K_{ij}$, $K$, and $^{(2)}R$ correspond to extrinsic, mean, and scalar curvature, respectively, and $a_{i}$ is a parameter related to the lapse function $N$.
%via $a_{i}=\partial_{i}\ln{N}$, being the line element in the preferred foliation
%\begin{equation}
%    ds^2=N^2dT^2-g_{ij}(dx^i+N^idT)(dx^j+N^jdT)\,.
%\end{equation}
Also, $g$ is the determinant of the induced metric,
%$g_{ij}$ on the constant-$T$ hypersurfaces.
 $L_4$ corresponds to a set of all the terms with four spatial derivatives that are invariant under diffeomorphisms. For $\lambda=\xi=1$ and $\eta=0$,  the action reduces to that of GR.
The covariantized reduced action of Ho\v{r}ava gravity is given by \cite{Sotiriou:2014gna}
\begin{equation}
    S_r=\frac{1}{8G_H} \int dt dr L_r\,,
\end{equation}
where
\begin{eqnarray}
\nonumber L_r&=&\frac{r^3F}{2Z}(\Omega^{\prime})^2-2\xi Z(\Lambda\frac{r}{F}+F^{\prime})+ \frac{r\eta F Z^{\prime 2}}{Z}+\frac{(1-\lambda)F^3 Z U^2}{r}+\frac{r F Z \left(1-\lambda +(1+\eta-\lambda)F^2U^2\right)}{1+F^2U^2}(UF^{\prime}+FU^{\prime})^2\\
&&+r(1+\eta-\lambda)F^2UZ^{\prime}\left (U(2F^{\prime}+\frac{FZ^{\prime}}{Z})+2FU^{\prime}\right)+2(\xi-\lambda)F^2U\left(FUZ^{\prime}+Z(UF^{\prime}+FU^{\prime})\right)\,,
\end{eqnarray}
%$G_{H}$ is a coupling constant with dimensions of a length squared and
 $Z(r)$, $F(r)$ and $\Omega(r)$ are the metric functions of a stationary and circularly symmetric metric given by
\begin{equation}\label{metric2}
ds^{2}=-Z(r)^2dt^{2}+\frac{1}{F(r)^2}dr^{2}+r^{2}(d\phi+\Omega(r)dt)^2~.
\end{equation}
In the sector $\eta=0$, Ho\v{r}ava gravity admits asymptotically AdS solutions \cite{Sotiriou:2014gna}, given by
\begin{equation}
F(r)^2= Z(r)^2 =-M +\frac{\bar{J}^2}{4r^2}-\bar{\Lambda}r^2~,
\end{equation}
with
\begin{equation}
\bar{J}^2=\frac{J^2+4a^2(1-\xi)}{\xi}~,~\Omega(r)=-\frac{J}{2 r^2}~,~\bar{\Lambda}=\Lambda-\frac{b^2(2\lambda -\xi-1)}{\xi}~,
\end{equation}
and
\begin{equation}
    U=\frac{1}{F}\left(\frac{a}{r}+br\right)\,.
\end{equation}
The aether configuration for this metric is
\begin{equation}
    u_r=\frac{1}{F^2}\left(\frac{a}{r}+br\right)\,,
\end{equation}
\begin{equation}
\label{ut}
    u_t=\sqrt{F^2+\left(\frac{a}{r}+br\right)^2}\,,
\end{equation}
where $a$ and $b$ are constants that can be regarded as measures of aether misalignment, with $b$ as a measure of asymptotically misalignment. For $b\neq 0$, the aether does not align with the timelike Killing vector asymptotically. Note that when $\xi=1$ and $\lambda=1$, the solutions become the BTZ black holes, and for $\xi=1$, the solutions become the BTZ black holes with a shifted cosmological constant given by $\bar{\Lambda}=\Lambda- 2b^2(\lambda-1)$. However, there is still a preferred direction represented by the aether vector
field which breaks Lorentz invariance for $\lambda \neq 1$ and $b\neq 0$. Also, $\bar{J}^2$ can be negative, when either $\xi < 0$ or $\xi > 1$, $a^2 > J^2/(4(\xi-1))$. The sign of $\bar{\Lambda}$  determine the asymptotic behavior (flat, dS, or AdS) of the metric \cite{Sotiriou:2014gna}. Here we only focus on the AdS sector.
The locations of the inner and outer horizons $r = r_\pm$, are given by
\begin{equation}\label{horizon}\
r_{\pm}^2=-\frac{M}{2\bar{\Lambda}}\left(1\pm \sqrt{1+\frac{\bar{J}^2\bar{\Lambda}}{M^2}} \right)~.
\end{equation}
Also, $M$ and $\bar{J}$ can be written in terms of the horizons as $M=-\bar{\Lambda}(r_+^2+r_-^2)$ and $\bar{J}= 2r_+ r_-\sqrt{-\bar{\Lambda}}$,
respectively. The Hawking temperature $T_H$ is given by $T_H=\frac{-\bar{\Lambda}(r_+^2-r_-^2)}{2\pi r_+}$.
For $\bar{J}\ne J$ (or equivalently $\xi\ne 1$), there is a curvature singularity due to the Ricci scalar $R=-6\bar{\Lambda}+\frac{1}{2 r^2}\left(\bar{J}^2-J^2\right)$ is divergent at $r=0$. This is in contrast to the BTZ black holes where the Ricci and Kretschmann scalars are finite and smooth at $r=0$.
Further notice that in these spacetimes there are different regions in the parameter space where there are no black hole solutions, the black holes have two horizons (inner and outer) and the black holes have only one horizon. In Fig. (\ref{co}), the different regions were defined by $\xi_e$, $\xi_c$, and $\bar{\Lambda}$ for a choice of the parameters \cite{Becar:2019hwk}.

\begin{figure}[!h]
\begin{center}
\includegraphics[width=70mm]{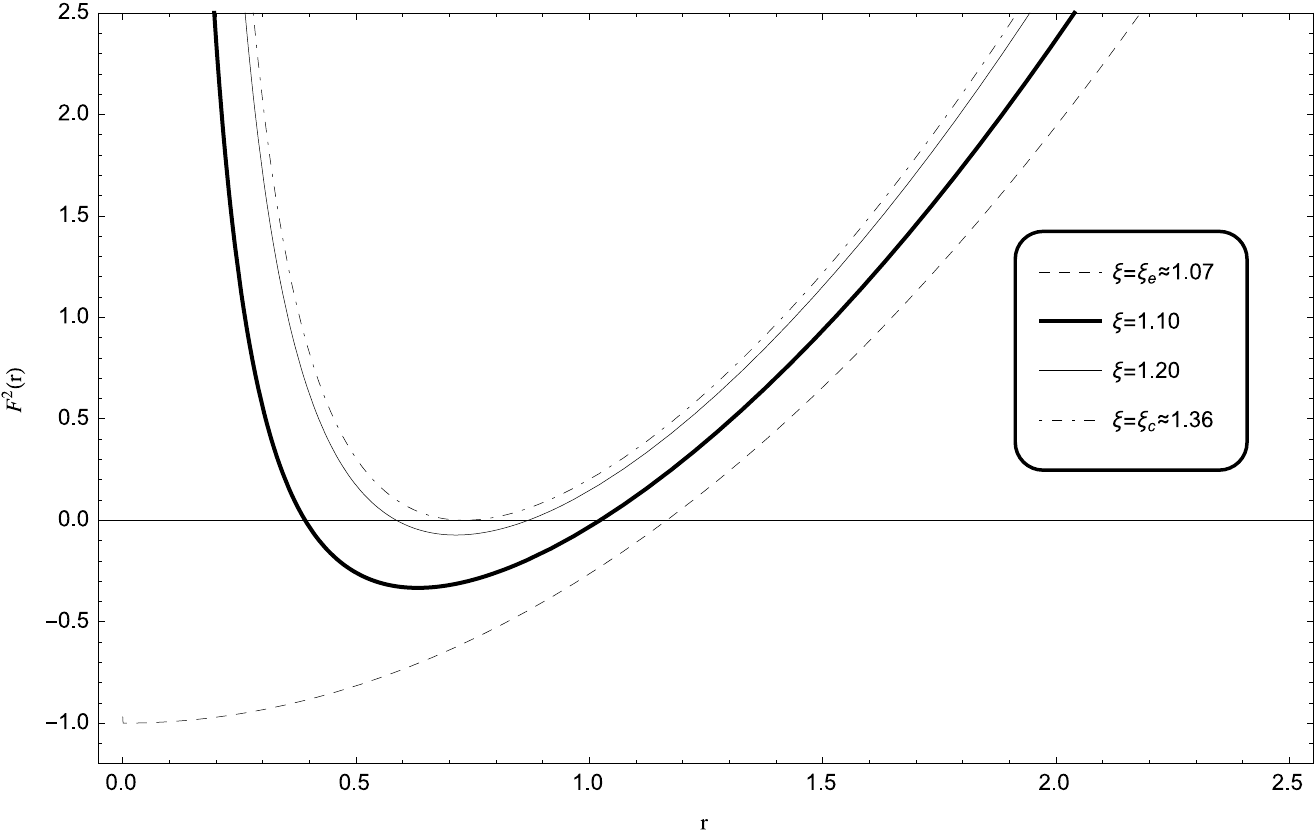}
\end{center}
\caption{The behavior of the lapsus function $F(r)^2=\mathcal{F}(r)$ as a function of $r$, with $M=1$, $\lambda=1$, $a=1$, $b=1$, $\Lambda=-1$, $J=1.2$, for different values of $\xi$. }
\label{fl}
\end{figure}

The value of $\xi$ for which the black hole is extremal is given by
\begin{eqnarray}
\notag  \xi_e &=&   -\frac{1}{2(M^2-4a^2 (b^2+ \Lambda))} \Big( b^2 (J^2+8a^2 \lambda)+ \Lambda (J^2+ 4 a^2) -\Big( (b^2 (J^2+8a^2 \lambda)+ \Lambda (J^2+ 4 a^2))^2   \\
 &&  +4b^2 (J^2+4 a^2)(2 \lambda-1) (M^2-4 a^2 (b^2 + \Lambda) )\Big)^{1/2}  \Big) \, ,
\end{eqnarray}
the value of $\xi$ for which the black hole goes from having two horizons to having one horizon is given by  $\xi_c = \frac{4 a^2 +J^2}{4 a^2}$,
and the value of $\xi$ for which the effective cosmological constant $\bar{\Lambda}$ changes sign is $\xi=\frac{(2 \lambda-1) b^2}{b^2 + \lambda}$. In Fig. (\ref{fl}), we show the behavior of the lapsus function $F(r)^2$, for a choice of parameters, and different values of $\xi$, where we can observe the existence of one horizon for $\xi=\xi_c\approx 1.36$, two horizons for $\xi_e<\xi<\xi_c$, and one horizon for $\xi=\xi_e\approx 1.07$.

\section{Equations of motion}
\label{ME}

The geodesic equations can be obtained from the Lagrangian of a test particle, which is given by \cite{chandra}
\begin{equation}
\mathcal{L}=\frac{1}{2}\left(g_{\mu\nu}\frac{dx^\mu}{d\tau}\frac{dx^\nu}{d\tau} \right)~.
\end{equation}
So, for the three-dimensional rotating Ho\v{r}ava AdS black hole described by the metric (\ref{metric2}), the Lagrangian associated with the motion of the test particles is given by
\begin{equation}\label{tl4}
2\mathcal{L}=-[\mathcal{F}(r)-r^2 \Omega^2(r)]\dot{t}^2+2r^2 \Omega(r)\,\dot{t}\,\dot{\phi}+
{ \dot{r}^2\over \mathcal{F}(r)}+ r^2\,\dot{\phi}^2~,
\end{equation}
where $\dot{q}=dq/d\tau$, and $\tau$ is an affine parameter along the geodesic. Here, we have defined $F(r)^2=Z(r)^2=\mathcal{F}(r)$.
%that
%we choose as the proper time.
Since the Lagrangian (\ref{tl4}) is
independent of the cyclic coordinates ($t,\phi$), then their
conjugate momenta ($\Pi_t, \Pi_{\phi}$) are conserved. Then, the equations of motion are obtained from
$ \dot{\Pi}_{q} - \frac{\partial \mathcal{L}}{\partial q} = 0$, and yield
\begin{equation}
\dot{\Pi}_{t} =0\,, \quad \dot{\Pi}_{r} =
-[\mathcal{F}'(r)/2-r \Omega^2(r)-r^2 \Omega'(r)]\dot{t}^{2}-
{\mathcal{F}'(r)\,\dot{r}^{2}\over 2\mathcal{F}^2(r)}
+r\,\dot{\phi}^2\,,\quad \textrm{and}\quad \dot{\Pi}_{\phi}=0~,
\label{w.11a}
\end{equation}
where $\Pi_{q} = \partial \mathcal{L}/\partial \dot{q}$
are the conjugate momenta to the coordinate $q$, and are given by
\begin{equation}
\Pi_{t} =-[\mathcal{F}(r)-r^2 \Omega^2(r)] \dot{t} +r^2 \Omega(r)\,\dot{\phi}\equiv -E~, \quad \Pi_{r}=
{ \dot{r}\over \mathcal{F}(r)}~\textrm{and}\quad \Pi_{\phi}
=r^2 \Omega(r) \dot{t} +r^2\,\dot{\phi}\equiv L~,
\label{w.11c}
\end{equation}
where $E$ and $L$ are integration constants associated to each of them.
Therefore, the Hamiltonian is given by
\begin{equation}
\mathcal{H}=\Pi_{t} \dot{t} + \Pi_{\phi}\dot{\phi}+\Pi_{r}\dot{r}
-\mathcal{L}\,.
\end{equation}
Thus,
\begin{equation}
2\mathcal{H}=-E\, \dot{t} + L\,\dot{\phi}+{ \dot{r}\over \mathcal{F}(r)}\equiv -m^2~,
\label{w.11z}
\end{equation}
where $m=1$ for timelike geodesics or $m=0$ for null geodesics.
Therefore, we obtain
\begin{eqnarray}
\label{w.14}
&&\dot{\phi}= -{1 \over (r^2-r_+^2)(r^2-r_-^2)\bar{\Lambda}}\left[ {EJ\over 2}
+L\left( {-\bar{\Lambda}r^2}
-M-{J^2-\bar{J}^2\over 4r^2}\right) \right] ~,\\
\label{w.12}
&&\dot{t}= -{ [E r^2-JL/2]\over (r^2-r_+^2)(r^2-r_-^2)\bar{\Lambda}}~,\\
\label{w.13}
&&\dot{r}^{2}=   \left( E-\frac{JL}{ 2\,r^2}\right)^2-\left( -M+\frac{\bar{J}^{2}}{4\, r^2}-\bar{\Lambda}r^2\right)  \left(m^2+\frac{L^2}{r^2}\right)=\left(E-V_-\right)\left(E-V_+\right)~,
\end{eqnarray}
where $V_{\pm}(r)$ is the effective potential and it is given by
\begin{equation}\label{tl8}
V_{\pm}(r)=\frac{JL}{ 2\,r^2}\pm\sqrt{\left( -M+\frac{\bar{J}^{2}}{4\, r^2}-\bar{\Lambda}r^2\right)  \left(m^2+\frac{L^2}{r^2}\right)}~.
\end{equation}
Since the negative branches have no classical interpretation, they are associated with antiparticles in the framework of quantum field theory \cite{Deruelle:1974zy}, we choose the positive branch of the effective potential $V = V_+$.
In the next section we will perform a general analysis of the equations of motion.

	\subsection{Dragging of inertial frames}
	In  a three-dimensional rotating Ho\v{r}ava AdS black hole, the presence of $g_{t\phi}\neq 0$
	in the metric introduces qualitatively new effects on test particles,
	like the effect  called the ``dragging of inertial frames''.
	Using Eqs. (\ref{w.14}) and (\ref{w.12}), we obtain	
	\begin{equation}
	\dot{\phi}/ \dot{t}={d\phi \over dt}={EJ/2-L[\bar{\Lambda}r^2+M+(J^2-\bar{J}^2)/4r^2]
		 \over Er^2-JL/2}
	\equiv \omega (r)\,.
	\end{equation}	
	This equation defines  the angular velocity of the test particle $\omega (r)$. Notice that, for vanishing angular momentum $L=0$	
	\begin{equation}
	\omega (r)={J \over 2\,r^2}\,.
	\end{equation}
	So, we have the remarkable result that a test particle dropped ``straight in'' from a finite distance is  ``dragged'' just by the influence of gravity so that it acquires an angular velocity  ($\omega$) with the same sign that of the source of the metric ($J$). This effect weakens with the distance as $1/r^2$, like in the BTZ metric \cite{Cruz:1994ir}.
	
		\subsection{Ergoregion}
	Following a similar treatment given in Ref. \cite{Shutz} for the Kerr geometry, we consider photons emitted  at some given $r$, initially going in the $\pm\phi$-direction, that is, tangent to a circle of constant $r$, $(dr=0)$. Then,
	from the metric (\ref{metric2}), we have
	\begin{equation}
	0=g_{tt} \,dt^2-J\,dt \, d\phi+r^2 \,d\phi^2\,,
	\end{equation}
	where $g_{tt}=M+\bar{\Lambda}r^2+{J^2-\bar{J}^2\over 4\,r^2}$. So, by using
	 $\omega =d\phi/dt$, we obtain
	\begin{equation}
	0=g_{tt} -J \, \omega+r^2\,\omega^2\,.
	\end{equation}
	Now, a remarkable effect happens if $g_{tt}=0$, the two solutions are
	\begin{equation}
	\omega_1=0 \quad \text{and} \quad \omega_2={J\over r^2}\,.
	\end{equation}
	The second solution, $\omega_2$,  gives  the same sign as the parameter $J$, and it represents the photon sent off in the same direction as the   rotating hole. The other solution, $\omega_1$,  means that the other photon the one sent  backwards initially does not move at all. The dragging of orbits has become so strong that this photon cannot move in the direction opposite to the rotation.
	The surface where $g_{tt}=0$  occurs at
	\begin{equation}
	r_{erg }=\left( -M/2\bar{\Lambda}+\sqrt{M^2/4\bar{\Lambda}^2-(J^2-\bar{J}^2)/4\bar{\Lambda}}\right) ^{1/2}\,.
	\end{equation}
	Inside this radius, since $g_{tt}>0$, all particles and photons must rotate with the black hole.
	The surface  lies outside the horizon; it is called the ergosphere. It is sometimes also called the ``static limit'', since inside it no particle can remain at fixed $ r,\phi$.  The result is different for the BTZ metric, where $r_{erg}^{BTZ}=(-M/\bar{\Lambda})^{1/2} < r_{erg}$ \cite{Banados:1992wn}.

%\newpage

\section{Null geodesics}
\label{geodesic}
\begin{figure}[!h]
	\begin{center}
		\includegraphics[width=80mm]{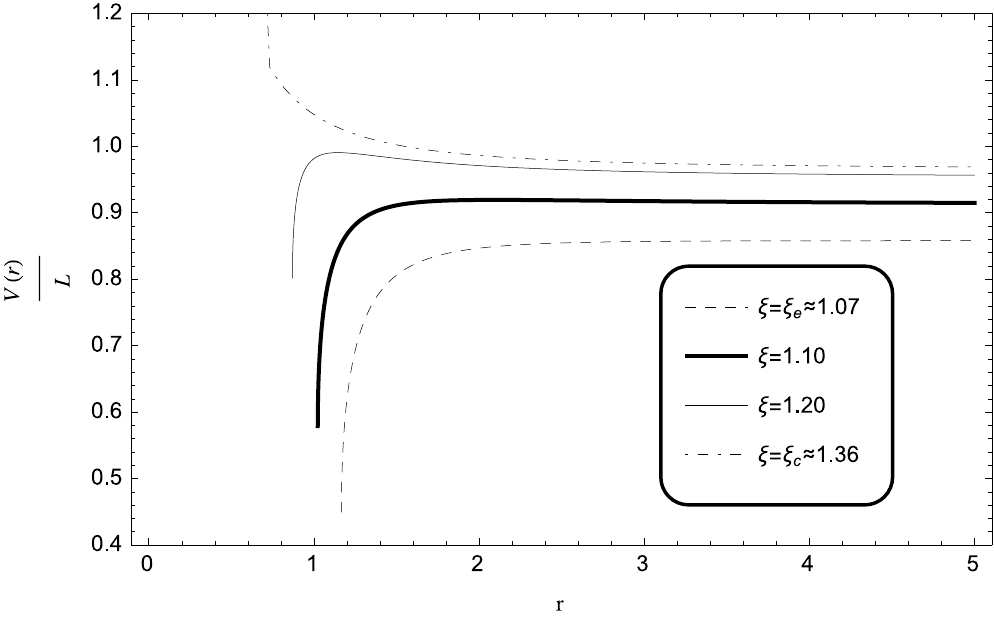}
			\includegraphics[width=80mm]{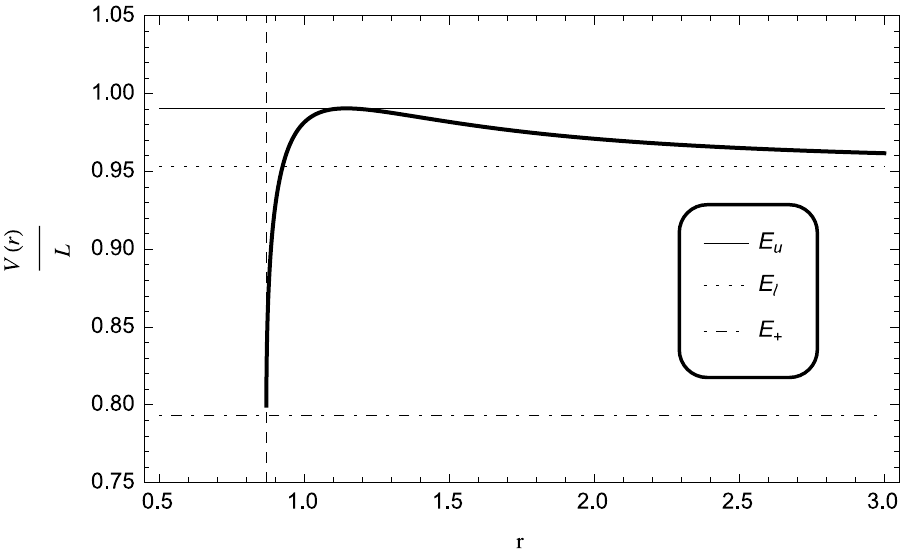}
	\end{center}
	\caption{The behavior of $V(r)/L$  as a function of $r$, for different values of $\xi$ (left panel) and for $\xi=1.1$ (right panel), with $M=a=b=\lambda=1$, $J=1.2$, $E_+=0.79$, $E_l=0.95$, $E_u=0.99$, and $\Lambda=-1$.}
	\label{f4.1}
\end{figure}

In this section we analyze the motion
of photons, $m^2=0$, so the
effective potential is given by
\begin{equation}
V\left( r\right) =\frac{JL}{ 2\,r^2}+\frac{L}{r}\sqrt{ -M+\frac{\bar{J}^{2}}{4\, r^2}-\bar{\Lambda}r^2 }~,
\label{t1}
\end{equation}
whose behavior is showed in Fig. \ref{f4.1} with
$J>0$,
where $E_{l}=L \sqrt{-\bar{\Lambda}}$ corresponds to the energy of the photon when $r\rightarrow \infty$, $E_+= V(r_+)$, and $E_u=V(r_
u)$, and $r_u$ is the radius at which the potential is maximum. We can distinguish two zones, one of them allows bounded orbits $(E\le E_u)$ and the other allows unbounded orbits $(E > E_u)$.

On the other hand, if we consider Eq. (\ref{w.14}) and Eq. ({\ref{w.13}}), the  orbit in polar coordinates is given by
\begin{equation}\label{e13}
 -{r^2\over (r^2-r_+^2)(r^2-r_-^2)\bar{\Lambda}}\left[ {EJ\over 2}
+L\left( {-\bar{\Lambda}r^2}
-M-{J^2-\bar{J}^2\over 4r^2}\right) \right]
\left(\pm \frac{dr}{d\phi}\right)= \sqrt{P(r)}\,.
\end{equation}
where $P(r)$ corresponds to the characteristic polynomial, and it is given by
\begin{equation}\label{tl12}
P(r)=r^4 \left(E^2+{L^2\bar{\Lambda}}\right)-L\left(JE-LM\right)r^2+(J^2-\bar{J}^2)L^2/4\,.
\end{equation}
Therefore, we can see that depending on the nature of its roots,
we can obtain the allowed motions for this spacetime.

\subsection{Radial motion}
\begin{figure}[!h]
	\begin{center}
		\includegraphics[width=80mm]{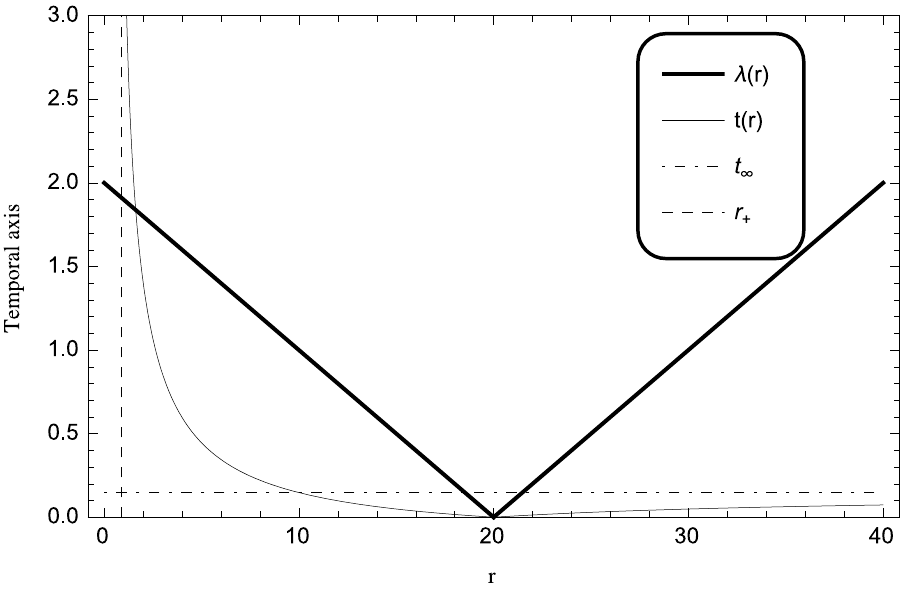}
			\end{center}
	\caption{The variation of the coordinate time $(t)$ and the affine parameter  $(\tau)$ along an unbounded time-like radial geodesic described by a photon as test particle, starting at $\rho_i=20$ and falling towards the singularity or going towards infinity, for $L=0$, $M=a=b=\lambda=1$, $\Lambda=-1$, $E=10$, $J=1.2$, $r_+\approx 0.87$, and $t_{\infty} \approx 0.147$.}
	\label{f444}
\end{figure}

\begin{figure}[!h]
	\begin{center}
	\includegraphics[width=90mm]{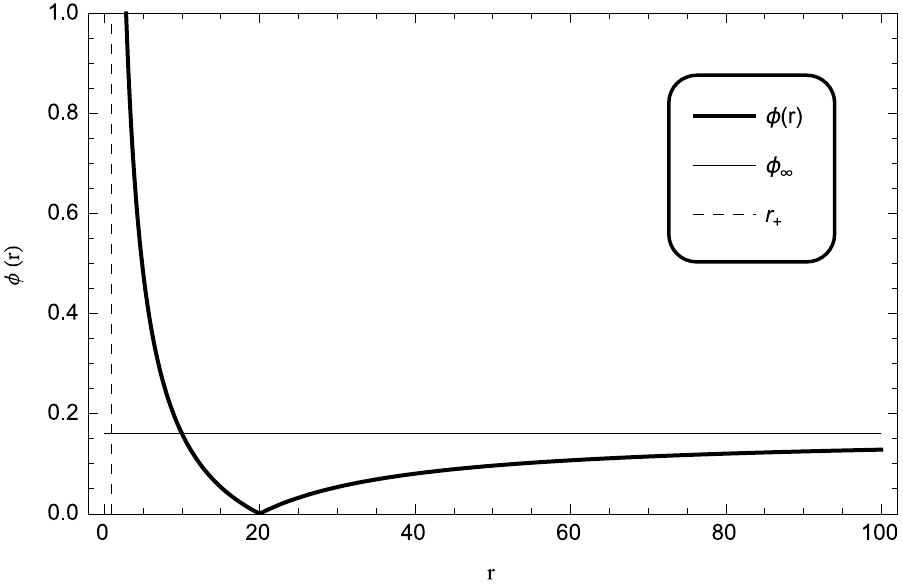}\hspace{5mm}
		\includegraphics[width=150mm]{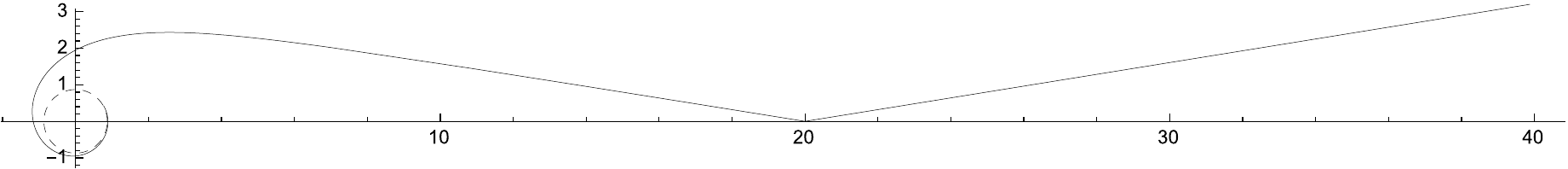}\hspace{5mm}
	\end{center}
	\caption{The behavior of the coordinate $\phi(r)$, starting at $\rho_i=20$ and falling towards the singularity or going towards infinity with  $L=0$, $M=a=b=\lambda=1$, $\Lambda=-1$, $J=1.2$, and $r_+=0.87$.
	In the polar plot, the trajectory approaching the horizon will spiral around the black hole an infinite number of times, and arrives at infinity in a finite coordinate $\phi$, i.e.,  $\phi \rightarrow \phi_ {\infty}\approx 0.16$ when $r \rightarrow \infty$.}
	\label{f4444}
\end{figure}

Radial motion corresponds to a trajectory with null angular
momentum $L=0$, and the photons are destined
to fall towards the event horizon or escape to infinity. From Eq.
(\ref{tl8}) we can see that for radial photons
$V(r)=0$, so that Eqs. (\ref{w.14}), (\ref{w.12}) and (\ref{w.13})  become
\begin{eqnarray}
\label{w.13r}
&&\dot{\phi}= -{E \,J\over 2(r^2-r_+^2)(r^2-r_-^2)\bar{\Lambda}}~,\\
\label{w.12r}
&&\dot{t}= -{E \,r^2\over (r^2-r_+^2)(r^2-r_-^2)\bar{\Lambda}}~,\\
\label{w.14r}
&&\pm \dot{r}=E~,
\end{eqnarray}
where the ($-$) sign corresponds to photons  falling into the event horizon and the ($+$) sign corresponds to photons that escape to infinity.
Choosing the initial conditions for the photons as $r=\rho_i$
when $\phi=t=\tau=0$, Eq. (\ref{w.14r}) yields
\begin{equation}
\tau(r)=\pm \frac{1}{E}(r-\rho_i)~,
\label{mr.3}
\end{equation}
where the $\pm$ signs have the same meaning given previously, and note that the above equation depends on the energy $E$. Also, for the negative sign the equation yields that the photons arrive to the event horizon in a finite affine parameter $\tau_+=(\rho_i-r_+)/E$, see Fig. \ref{f444}. On the other hand, a straightforward integration of Eqs. (\ref{w.12r}) and (\ref{w.13r}) leads to
\begin{equation}
t(r)=\mp {1\over 2\bar{\Lambda}(r^2_{+}-r^2_{-})}\left[r_{+}
\ln \left|\frac{r-r_{+}}{\rho_i-r_{+}}\cdot\frac{\rho_i+r_{+}}{r+r_{+}}\right|-
r_{-}
\ln \left|\frac{r+r_{-}}{\rho_i+r_{-}}\cdot\frac{\rho_i-r_{-}}{r-r_{-}}\right|
\right]~,
\label{mr.5}
\end{equation}
and
\begin{equation}
\phi(r)=\mp {J \over 4\bar{\Lambda}(r^2_{+}-r^2_{-})}\left[{1\over r_{+}}
\ln \left|\frac{r-r_{+}}{\rho_i-r_{+}}\cdot\frac{\rho_i+r_{+}}{r+r_{+}}\right|-
{1\over r_{-}}
\ln \left|\frac{r+r_{-}}{\rho_i+r_{-}}\cdot\frac{\rho_i-r_{-}}{r-r_{-}}\right|
\right]~.
\label{mr.5b}
\end{equation}
The solution for the coordinate time $t$ does not depend on the energy of the photon, and the solution for $\phi$, see Fig. \ref{f4444}, neither depend on the energy of the photon but it depends on the angular momentum parameter $J$. Moreover, for radial geodesics when $r \rightarrow \infty$ we find
\begin{equation}
t_{\infty}= - {1\over 2\bar{\Lambda}(r^2_{+}-r^2_{-})}\left[r_{+}
\ln \left|\frac{\rho_i+r_{+}}{\rho_i-r_{+}}\right|-
r_{-}
\ln \left|\frac{\rho_i-r_{-}}{\rho_i+r_{-}}\right|
\right]~,
\label{mr.5}
\end{equation}
and
\begin{equation}
\phi_{\infty}= -{J\over 4\bar{\Lambda}(r^2_{+}-r^2_{-})}\left[{1\over r_{+}}
\ln \left|\frac{\rho_i+r_{+}}{\rho_i-r_{+}}\right|-
{1\over r_{-}}
\ln \left|\frac{\rho_i-r_{-}}{\rho_i+r_{-}}\right|
\right]~.
\label{mr.5b}
\end{equation}

Thus, as seen by a system external to photons, they will fall asymptotically to the event horizon. On the other hand, this external observer will see that photons arrive in a finite coordinate time to spatial infinite. Remarkably, the behavior is similar to the observed in Lifshitz's spacetimes \cite{Villanueva:2013gra}.

\newpage

\subsection{Unstable circular orbit}

The unstable circular orbit is a particular orbit for the Ho\v{r}ava spacetime, which is not allowed in the BTZ spacetime. The effective potential
is maximum at

\begin{equation}
\label{ru}
r_u=\left[ -\frac{ M(J^2-\bar{J}^2)}{2\bar{\Lambda}(J^2+M^2/\bar{\Lambda})}+
\sqrt{\left( -\frac{ M(J^2-\bar{J}^2)}{2\bar{\Lambda}(J ^2+M^2/\bar{\Lambda})}\right)^2+ \frac{\bar{J}^2(J^2-\bar{J}^2)}{4\bar{\Lambda}(J ^2+M^2/\bar{\Lambda})}}\right]^{1/2}\,,
\end{equation}
where $J>M/\sqrt{-\bar{\Lambda}}>\bar{J}$. The energy for this orbit is given by $E_u=V(r_u)$. The period of a revolution according to the affine parameter, in the unstable circular orbit is
\begin{equation}\label{p1}
T_{\lambda}= - {4\pi r_u \bar{\Lambda} (r_u^2-r_+^2)(r_u^2-r_-^2) \over L(J\sqrt{F(r_u)}+2r_uF(r_u))}\,,
\end{equation}
and the period  according to the coordinate time, in the unstable circular orbit is
\begin{equation}\label{p1}
T_t={4\pi r_u ^2\sqrt{F(r_u)}\over J\sqrt{F(r_u)}+2r_uF(r_u)}\,.
\end{equation}
Thus, while the revolution period $T_{\lambda}$ depends on the horizons ($r_{\pm}$), the period respect to the coordinate time $T_t$ does not depend on the horizons.\\

\begin{figure}[!h]
\begin{center}
\includegraphics[width=85mm]{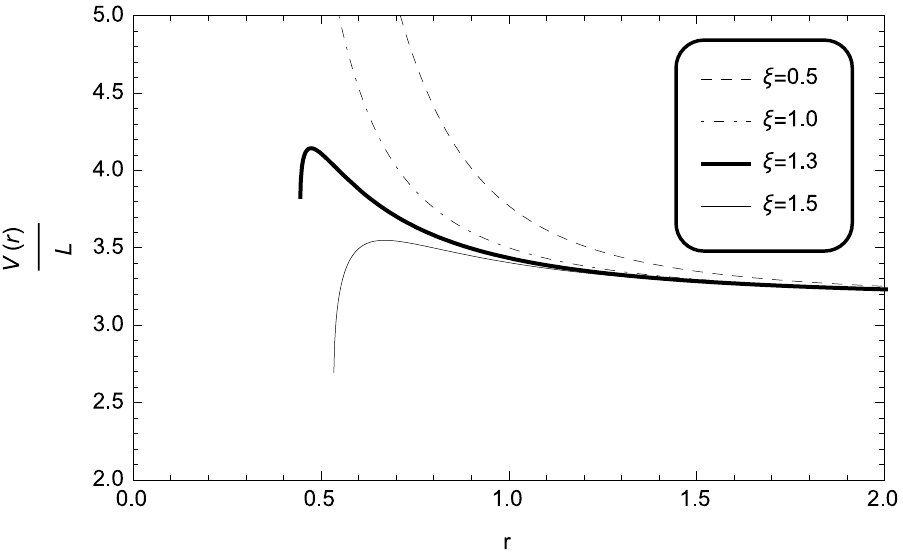}
\end{center}
\caption{The behavior of $V(r)/L$  as a function of $r$, for $a=b=0$, $\lambda=1$, $\Lambda=-10$, $J=1.5$, $M=3$ and different values of $\xi$. }
\label{figu1}
\end{figure}

\begin{figure}[!h]
\begin{center}
\includegraphics[width=100mm]{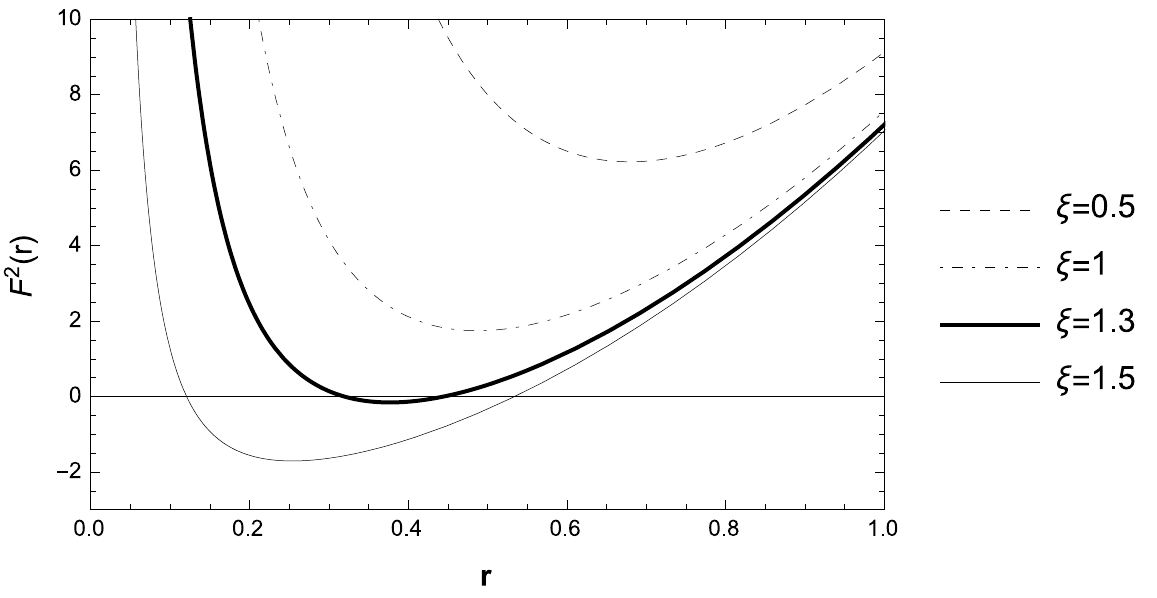}
\end{center}
\caption{The behavior of the lapsus function $F(r)^2=\mathcal{F}(r)$ as a function of $r$, for $a=b=0$, $\lambda=1$, $\Lambda=-10$, $J=1.5$, $M=3$ and different values of $\xi$. }
\label{figu2}
\end{figure}

Notice that the radius of the unstable circular orbit given by Eq. (\ref{ru}), when it exists, it can be written as:

\begin{equation}
r_{u}^2 = \frac{\xi (1- \xi) M (J^2+4 a^2)- \sqrt{\xi^2 (1- \xi) J^2 (J^2+4 a^2) (-J^2 \bar{\Lambda}-M^2 \xi - 4 \bar{\Lambda} a^2 (1- \xi))}}{2 \xi^2 (M^2+ J^2 \bar{\Lambda})}\,.
\end{equation}
We observe that for $\xi=1$ the above expression is null and the unstable circular orbit for photons does not exist. 
Now, we will analyze the behavior of the effective potential for a fixed value of $\lambda$ and different values of the parameter $\xi$ and for fixed values of the integration constants $M$, $J$, $a$ and $b$. For $\lambda=\xi=1$ the action of Ho\v{r}ava gravity in the infrared limit with $\eta=0$ reduces to the action of general relativity, while for $\xi \neq 1$ and/or $\lambda \neq 1$ the theory is not local Lorentz invariant. In this way, we can link the existence of unstable circular orbit to the breaking of the local Lorentz symmetry of the theory when $\xi \neq 1$. In the next, we will fix $\lambda=1$, and study only the effect of varying $\xi$. In the infrared limit Ho\v{r}ava gravity is equivalent to the Einstein-aether theory, and for $\xi=1$ (and $\lambda=1$) we recover the BTZ metric and the parameters $a$ and $b$, which are related to the aether misalignment, do not appear in the metric in this case. Also, in the following we set $b=0$, i.e. the aether is alignment with the timelike Killing vector asymptotically,  and to avoid the eather becoming imaginary for small $r$ we take $a \neq 0$, see Eq. (\ref{ut}); so, $\xi$ is the only parameter of the action that appears in the metric and in the effective potential. We focus in this case in order to see the effect of the parameter $\xi$, which is related to the breaking of the local Lorentz invariance of the theory, on the effective potential. In Fig. (\ref{figu1}) we observe that for $\xi=0.5,1$ the unstable circular orbit is not possible, while that for $\xi=1.3, 1.5$ there appear local maximum in the potential indicating the existence of unstable circular orbits. Fig. (\ref{figu2}) shows that $\xi=0.5, 1$ corresponds to naked singularities for the chosen values of the parameters, and the cases $\xi=1.3, 1.5$ represent black holes. It is worth mentioning that the unstable circular orbit also is possible in a family of three-dimensional asymptotically AdS static black holes in New Massive Gravity, where the existence of such geodesics is determined by the sign of the hair parameter \cite{Acena:2019jvf}.

\subsection{Critical trajectories}

\begin{figure}[!h]
	\begin{center}
		\includegraphics[width=80mm]{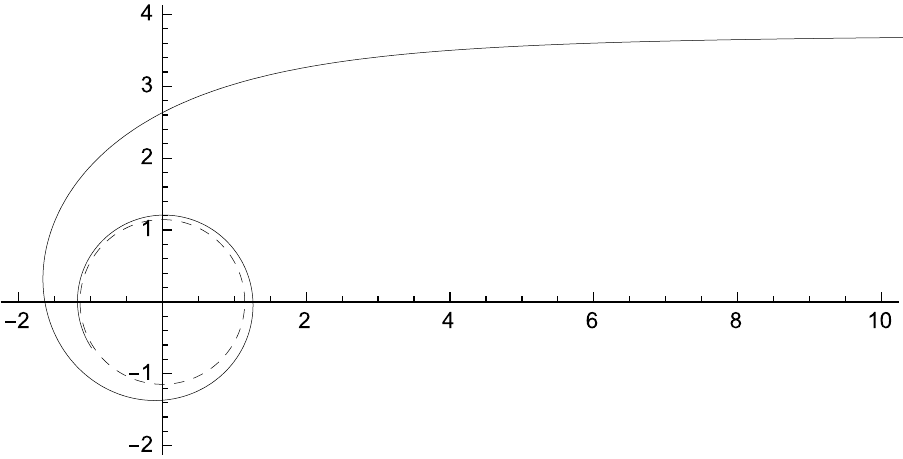}
	\end{center}
	\caption{Critical trajectory of first kind, for photons with  $L=1$, $M=a=b=\lambda=1$, $\Lambda=-1$, $J=1.2$, and $E_u=0.99$. For orbit of the first kind (thin line) the test photon arrived from infinity to $r_u= 1.15$, where $r_u$ corresponds to the radius of the unstable circular orbit (dashed circle).}
	\label{fC}
\end{figure}

\begin{figure}[!h]
	\begin{center}
		\includegraphics[width=40mm]{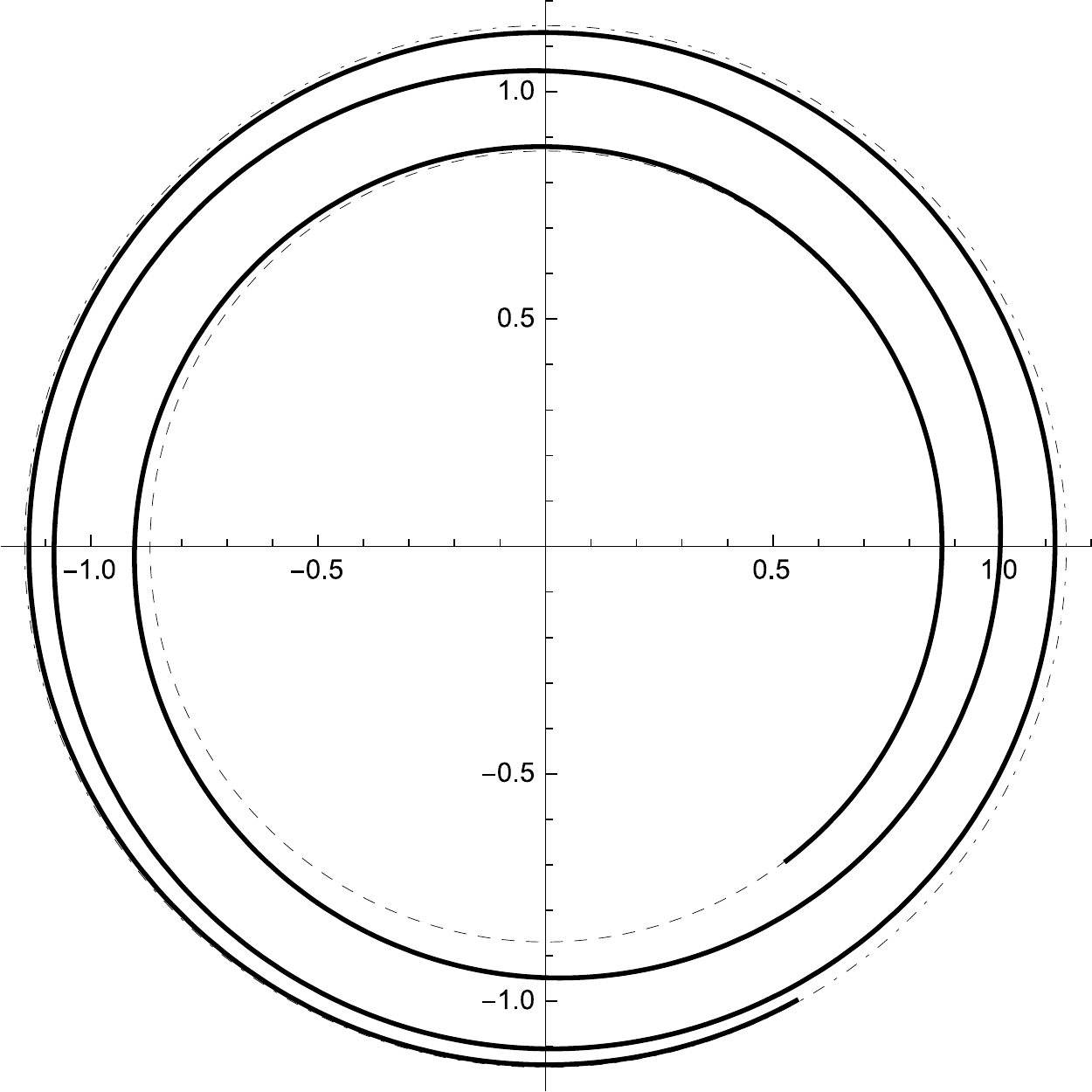}
	\end{center}
	\caption{Critical trajectory of second kind, for photons with  $L=1$, $M=a=b=\lambda=1$, $\Lambda=-1$, $J=1.2$, and $E_u=0.99$. For orbit of the second kind (black line), the trajectory is asymptotic to two circles $r_+= 0.87$ (dashed circle) and $r_u= 1.14$ (dot dashed circle).}
	\label{fC2}
\end{figure}

The unstable circular orbits of radius $r_u$  correspond to the maximum of the potential. In this case, the energy of the photon is $E_u = V(r_u)$. Also, there are two critical orbits that approach to the unstable circular orbit asymptotically. In the first kind, the particle arises from infinity, see Fig. \ref{fC}, and in the second kind, the particle starts from a finite distance bigger than the horizon radius, but smaller than the unstable radius, see Fig. \ref{fC2}. The solution for the critical orbit of first kind is

\begin{equation}
\phi_C(r)= k_0 \sum _{i=1}^3 k_i \Theta_i(r)\,,
\end{equation}
where
\begin{equation}
    k_0=-\frac{1}{2 \bar{\Lambda}  r_u^3 \left(r_+^2-r_c^2\right)\sqrt{E^2_u+\bar{\Lambda}L^2}}\,,
\end{equation}
\begin{equation}
k_1=\frac{2 r_+^3 r_u^3}{r_u^2-r_+^2} \left(\frac{LM-E_u J/2}{r_+^2}+\frac{L(J^2-\bar{J}^2)}{4 r_+^4}+\bar{\Lambda}L\right)\,,
\quad
k_2=-\frac{2 r_-^3 r_u^3}{r_u^2-r_-^2} \left(\frac{LM-E_u J/2}{r_-^2}+\frac{L(J^2-\bar{J}^2)}{4 r_-^4}+\bar{\Lambda}L\right)\,,
\end{equation}
\begin{equation}
k_3=-\frac{r_u^2 \left(r_+^2-r_-^2\right) }{\left(r_u^2-r_-^2\right) \left(r_u^2-r_+^2\right)} \left(r_u^2 \left(LM-E_u J/2\right)+\frac{L(J^2-\bar{J}^2)}{4}+\bar{\Lambda}L r_u^4\right)\,,
\end{equation}
and
\begin{equation}
\Theta_1 (r)=\tanh ^{-1}\left(\frac{r_e}{r}\right)\,,
\quad
\Theta_2 (r)=\tanh ^{-1}\left(\frac{r_c}{r}\right)\,,
\quad
\Theta_3 (r)=\ln \left(\frac{r+r_u}{r-r_u}\right)\,.
\end{equation}

The solution for the critical orbit of second kind is
\begin{equation}
\phi_C(r)= \tilde{k}_0 \sum _{i=1}^3 \tilde{k}_i \tilde{\Theta}_i(r)\,,
\end{equation}
where

\begin{equation}
\tilde{k}_0=k_0\,, \quad
\tilde{k}_1=k_1\,, \quad
\tilde{k}_2=k_2\,, \quad
\tilde{k}_3=-k_3\,,
\end{equation}
and
\begin{equation}
\tilde{\Theta}_1 (r)=\tanh ^{-1}\left(\frac{r_e}{r}\right)-\tanh ^{-1}\left(\frac{r_e}{r_0}\right)\,,
\quad
\tilde{\Theta}_2 (r)=\tanh ^{-1}\left(\frac{r_c}{r}\right)-\tanh ^{-1}\left(\frac{r_c}{r_0}\right)\,,
\end{equation}
\begin{equation}
\tilde{\Theta}_3 (r)=\ln \left(\frac{r_u-r}{r_u-r_0}\cdot\frac{r_u+r_0}{r_u+r}\right)\,.
\end{equation}

\newpage

\subsection{Deflection of light}
Orbits of the first kind occur when the energy $E$ lies in the range $E_{\ell}<E<E_u$,
and this case requires that $P(r)=0$ allows three real roots, which we can identify as  $r_d$, which correspond to the closest distance,  $r_f$ as an apoastro distance for the trajectories of the second kind.
%and the third  root, $r_3$ and $r_4$ are negative without physical interest.
Thus, we can rewrite
the characteristic polynomial (\ref{tl12}) as
\begin{equation}\label{c10}
P(r)= \left(E^2-E^2_{\ell}\right)(r^2-r^2_d)(r^2-r^2_f),
\end{equation}
where
\begin{equation}
r_d=\left[ \frac{J L E -L^2 M}{2(E ^2+\bar{\Lambda}L^2)}+\sqrt{\left(\frac{J L E -L^2 M}{2(E ^2+\bar{\Lambda}L^2)}\right)^2-\frac{L^2 \left(J^2-\bar{J}^2\right)}{4(E ^2+\bar{\Lambda}L^2)}}\right] ^{1/2}~,
\label{rd}
\end{equation}
and
\begin{equation}
r_f=\left[ \frac{J L E -L^2 M}{2(E ^2+\bar{\Lambda}L^2)}-\sqrt{\left(\frac{J L E -L^2 M}{2(E ^2+\bar{\Lambda}L^2)}\right)^2-\frac{L^2 \left(J^2-\bar{J}^2\right)}{4(E ^2+\bar{\Lambda}L^2)}}\right] ^{1/2}~.
\label{rf}
\end{equation}
So, choosing the initial conditions for the photons as $r=r_d$
when $\phi=t=\tau=0$, Eqs. (\ref{w.14}) and (\ref{w.12}) yield
\begin{equation}
\phi(r)=\int_{r_d}^{r} - {r^2\over \bar{\Lambda}(r^2-r_+^2)(r^2-r_-^2)}\left[
-L\bar{\Lambda}r^2+\left( {EJ\over 2}
-ML\right) -L{J^2-\bar{J}^2\over 4r^2}\right]{d\,r\over \sqrt{P(r)}}~.
\end{equation}

The solution for $\phi(r)$ is
\begin{equation}
\phi(r)=\delta_0\sum^{3}_{j=1}\delta_j\,\left( \varPhi_j[U(r)]- \varPhi_j[U(r_d)]\right) ~,
\label{phi1}
\end{equation}
where
\begin{equation}
    \delta_0=-\frac{r_d^2}{\bar{\Lambda}L(r_+^2-r_-^2)\sqrt{J^2-\bar{J}^2}}\,,
    \quad
    \delta_1 =\frac{L(J^2-\bar{J}^2)}{4r_d^3}\,,
\end{equation}
\begin{equation}
    \delta_2 =\frac{1}{4}\left[\frac{r_d}{r_+^2}(\frac{EJ}{2}-LM)-\bar{\Lambda}Lr_d-\frac{Lr_d}{r_+^4}\frac{(J^2-\bar{J}^2)}{4}\right]\,,
    \quad
    \delta_3 =-\frac{1}{4}\left[\frac{r_d}{r_-^2}(\frac{EJ}{2}-LM)-\bar{\Lambda}Lr_d-\frac{Lr_d}{r_-^4}\frac{(J^2-\bar{J}^2)}{4}\right]\,,
\end{equation}
and
\begin{equation}
\varPhi_1(r)=\wp^{-1}[U(r)]\,,
\end{equation}
\begin{equation}
\varPhi_2(r)=\frac{1}{\wp^{'}(\Omega_2)}
\left[\zeta(\Omega_2)\wp^{-1}[U(r)]
+\ln\left|\frac{\sigma[\wp^{-1}[U(r)]-\Omega_2]}
{\sigma[\wp^{-1}[U(r)]+\Omega_2]}\right|
\right]\,,
\end{equation}
\begin{equation}
\varPhi_3(r)=\frac{1}{\wp^{'}(\Omega_3)}
\left[\zeta(\Omega_3)\wp^{-1}[U(r)]
+\ln\left|\frac{\sigma[\wp^{-1}[U(r)]-\Omega_3]}
{\sigma[\wp^{-1}[U(r)]+\Omega_3]}\right|
\right]\,,
\end{equation}
where
\begin{equation}
U(r)={r_d^2\over 4r^2}+\frac{r_d^2}{3L}\left(\frac{LM-JE}{J^2-\bar{J}^2}\right)\,,
\end{equation}
\begin{equation}
\Omega_2=\wp^{-1}\left[\frac{r_d^2}{4r_+^2}+\frac{r_d^2(M-JE/L)}{3(J^2-\bar{J}^2)}\right]\,,
\quad
\Omega_3=\wp^{-1}\left[\frac{r_d^2}{4r_-^2}+\frac{r_d^2(M-JE/L)}{3(J^2-\bar{J}^2)}\right]
\,,
\end{equation}
\begin{equation}
g_2=\frac{r_d^4}{L^2}\left[\frac{4}{3}\left(\frac{LM-JE}{J^2-\bar{J}^2} \right)^2-\frac{E^2-E_l^2}{J^2-\bar{J}^2}\right]\,,
\quad
g_3=\frac{r_d^6}{L^3}\left[\frac{(E^2-E_l^2)(LM-JE)}{3(J^2-\bar{J}^2)^2} -\frac{8}{27}\left(\frac{LM-JE}{J^2-\bar{J}^2}\right)^3\right]\,.
\end{equation}

\begin{figure}[!h]
	\begin{center}
		\includegraphics[width=40mm]{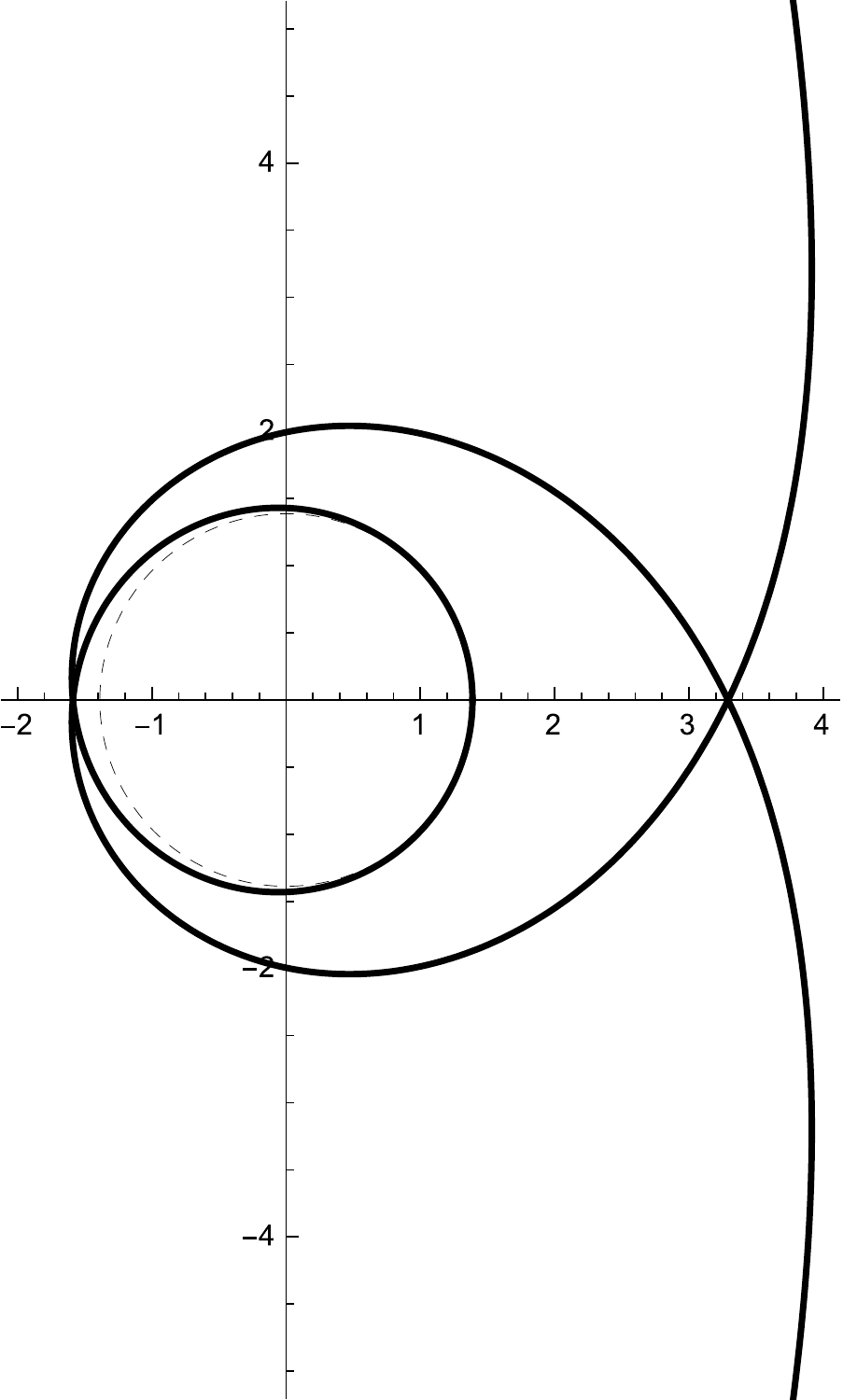}
	\end{center}
	\caption{The behavior of the deflection of light with $E=0,98$,  $L=1$, $M=a=b=\lambda=1$, $\Lambda=-1$, $J=1.2$. Here,  the photon starts its trajectory from infinity. Then, it approaches to the black hole until the closest distance $r_d=1.4$ (dashed line), and finally returns to infinity.}
	\label{f8}
\end{figure}

\subparagraph*{The deflection angle. }
	As an application of the preceding  solution, specifically from the polar equation (\ref{phi1}),
	%for the photon orbit of the first kind,
	it is possible to study how the photon is deflected due to presence  of a three-dimensional rotating Ho\v{r}ava AdS black hole, whose singularity, for  simplicity, is placed at the origin of a coordinate system.  The deflection angle $\hat{\alpha}=2|\phi_{\infty}|-\pi$, 	where $\phi_{\infty}=\phi(\infty)$, is given by
		\begin{equation}
\hat{\alpha}=2\left| \delta_0\sum^{3}_{j=1}\delta_j\,
\left( \varPhi_j\left[ \frac{r_d^2}{3L}\left(\frac{LM-JE}{J^2-\bar{J}^2}\right)\right] - \varPhi_j\left[\frac{1}{4}+ \frac{r_d^2}{3L}\left(+\frac{LM-JE}{J^2-\bar{J}^2}\right)\right] \right) \right| -\pi\,.
	\end{equation}
	This is an exact solution for the angle of deflection, and it depends on the spacetime parameters; $M,\, J, \,\bar{J}$, and the photon motion constants, $E$ and $L$. In Fig. \ref{f8}, we plot the behavior of the deflection of light in the background of a three-dimensional rotating Ho\v{r}ava AdS black hole.

%%%%%%%%%%%%%%%%%%%%%%%%%%%%%%%%%%%%%%%%%%%%%%%%%%%%%%%%%%%%%
%%%%%%%%%%%%%%%%%%%%%%%%%%%%%5

%%%%%%%%%%%%%%%%%%%%%%%%%%%%%%%%%

%\newpage

\subsection{Second kind trajectory}

\begin{figure}[!h]
	\begin{center}
		\includegraphics[width=40mm]{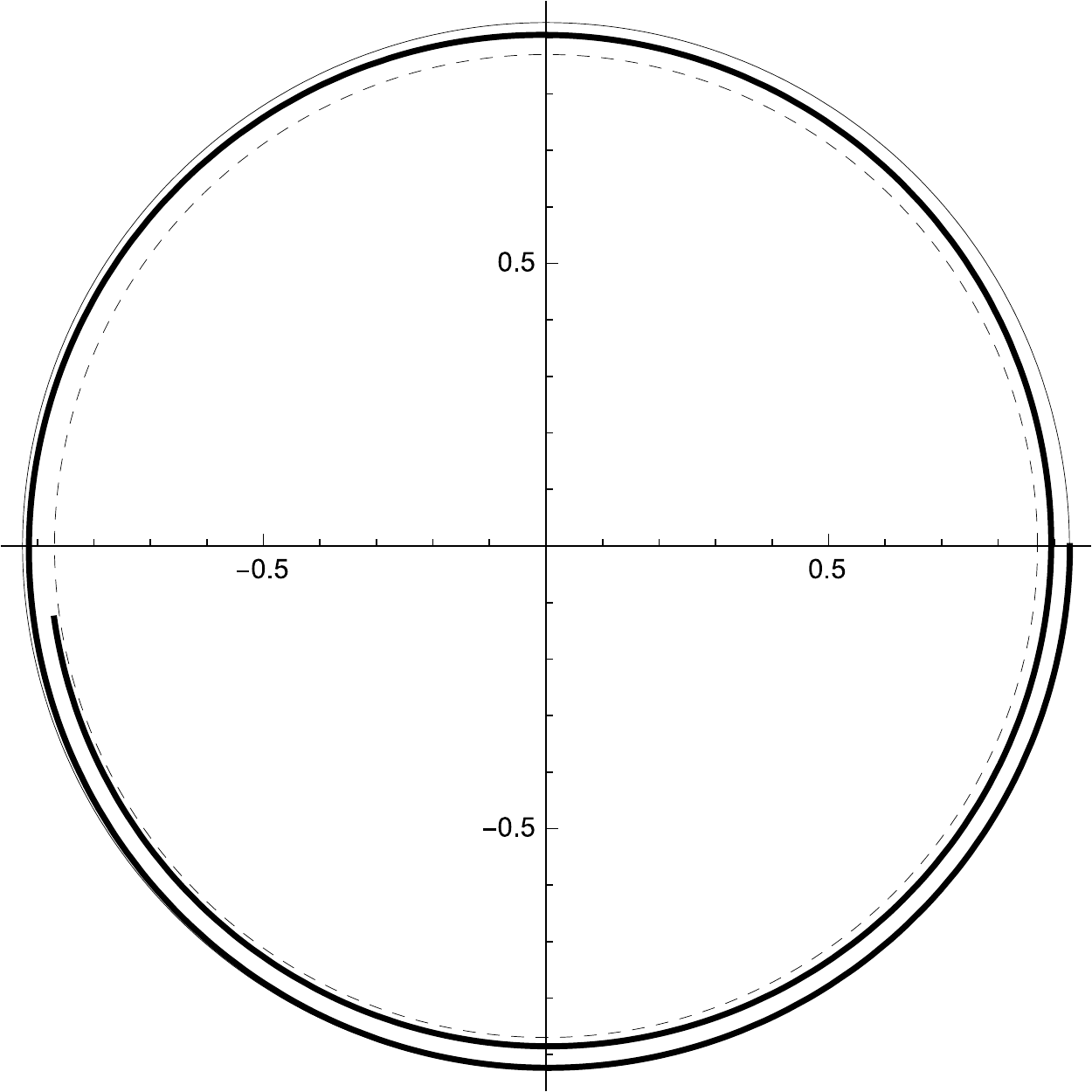}
	\end{center}
	\caption{This trajectory corresponds to the one where the test photon has a return point at $ r_{\ell}$,  bigger than the event horizon and then falls to it, asymptotically, by spiralling around it. Its motion is developed in the region  $r_+< r < r_{\ell}$, with $E_l=0.95$,  $L=1$, $M=a=b=\lambda=1$, $\Lambda=-1$, $J=1.2$, $r_{\ell}=0.93$ and $r_+=0.87$.}
	\label{f9}
\end{figure}

Orbits of the second kind corresponds to trajectories with a return point from which plunges to the event horizon, see Fig. \ref{f9}. These orbits are allowed in the energy range $E_{+}<E<E_u$. The return point is in the range $r_+<r<r_u$. By simplicity, we consider the trajectory with $E=E_l$ with a return point $r=r_l$, where $E_l=V(\infty)$. Thus, we can rewrite
the characteristic polynomial (\ref{tl12}) as
\begin{equation}\label{c10}
P(r)=L^2\left(J/\ell-M \right)  (r^2_{\ell}-r^2)\,,
\end{equation}
where
\begin{equation}
r_{\ell}={1\over 2}\sqrt{\frac{J^2-\bar{J}^2}{J/\ell-M}}\,.
\label{rd}
\end{equation}
So, choosing the initial conditions for the photons as $r=r_d$
when $\phi=t=\tau=0$, Eqs. (\ref{w.14}) (\ref{w.12})  yield the following solution
\begin{equation}
\phi(r)=\eta_0\sum^{3}_{j=1}\eta_j\,\Psi_j[r]  ~,
\label{phi11}
\end{equation}
where
\begin{equation}
\eta_0={\ell^2\over L(r_+^2-r_-^2)(J^2-\bar{J}^2)^{1/2}}\,,
\quad
\eta_1 ={ 2Lr_{\ell}^2(r_+^2-r_-^2) \over \ell^2}\,,
\end{equation}
\begin{equation}
\eta_2 = \frac{r_+}{\sqrt{r_{\ell}^2-r_+^2}}\left[ r_{\ell}^2\left(\frac{EJ}{2}-LM\right)+\frac{Lr_{\ell}^2r_+^2}{\ell^2}-\frac{Lr_{\ell}^2(J^2-\bar{J}^2)}{4r_+^2}\right]\,,
\end{equation}
\begin{equation}
\eta_3 = - \frac{r_-}{\sqrt{r_{\ell}^2-r_-^2}}\left[ r_{\ell}^2\left(\frac{EJ}{2}-LM\right)+\frac{Lr_{\ell}^2r_-^2}{\ell^2}-\frac{Lr_{\ell}^2(J^2-\bar{J}^2)}{4r_-^2}\right]\,,
\end{equation}
and
\begin{equation}
\Psi_1(r)=\tan^{-1}\sqrt{{r_{\ell}^2\over r^2}-1}\,,
\quad
\Psi_2(r)=\ln \left|  \frac{\sqrt{{r_{\ell}^2\over r^2}-1}+\sqrt{{r_{\ell}^2\over r_+^2}-1}}{\sqrt{{r_{\ell}^2\over r^2}-1}-\sqrt{{r_{\ell}^2\over r_+^2}-1}}        \right|\,,
\quad
\Psi_3(r)=\ln \left|  \frac{\sqrt{{r_{\ell}^2\over r^2}-1}+\sqrt{{r_{\ell}^2\over r_-^2}-1}}{\sqrt{{r_{\ell}^2\over r^2}-1}-\sqrt{{r_{\ell}^2\over r_-^2}-1}}        \right|\,.
\end{equation}

\newpage

\subsection{Unbounded trajectories}

\begin{figure}[!h]
	\begin{center}
		\includegraphics[width=70mm]{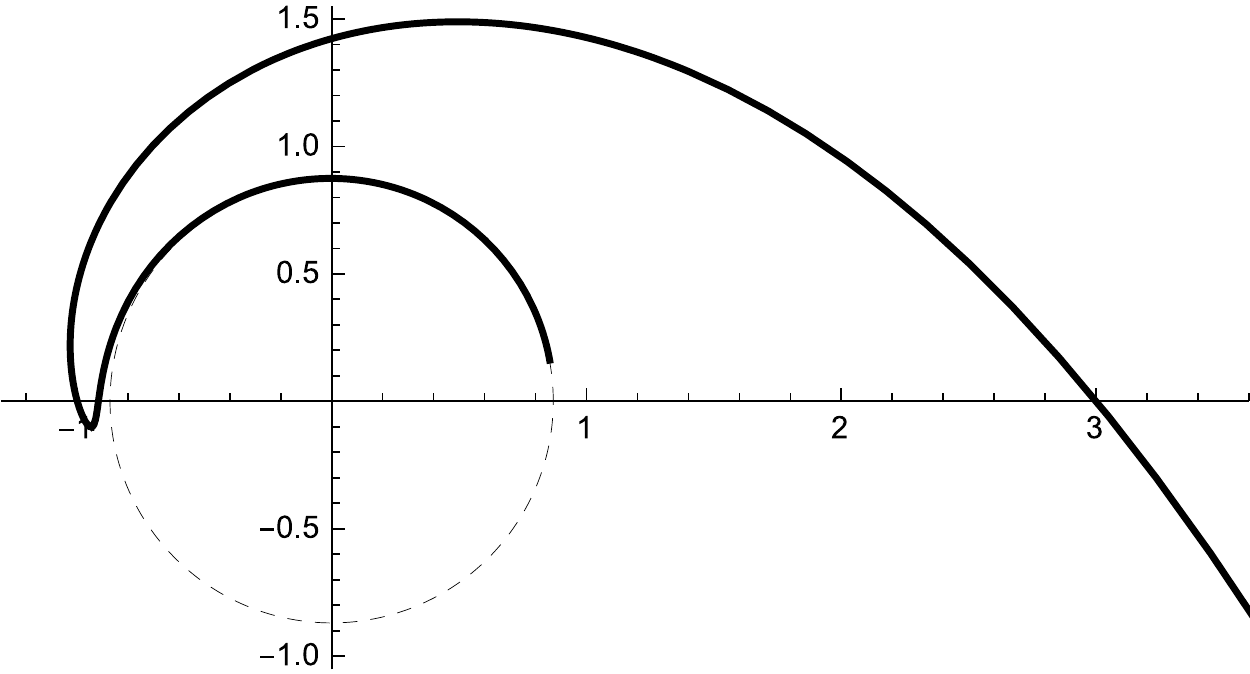}
	\end{center}
	\caption{Polar plot for an unbounded  trajectory. The photons arriving from infinity approach to the horizon (dashed line), asymptotically, by spiralling around it, with $E=1.03$,  $L=1$, $M=a=b=\lambda=1$, $\Lambda=-1$, $J=1.2$  and $r_+=0.87$. }
	\label{f10}
\end{figure}

Finally, we consider the region where the photons are captured, or escape to infinity, depending on initial conditions, and its cross section ($\sigma$) is $\sigma=\pi b_c^2$, where $b_c$ corresponds to the impact parameter given by $L/E_u$,  see Fig. \ref{f10}, which corresponds to an energy parameter $E> E_u$. The solution of the trajectory is similar to the solution of the deflection of the light, considering the corresponding range of energy and radial distance: $r_+<r<\infty$.

%\newpage

\section{Remarks and conclusions}
\label{conclusion}

In this work, we studied the motion of photons in the background of a rotating three-dimensional Ho\v{r}ava AdS black hole described by a Lorentz-violating version of the BTZ black hole,  i.e.  a black hole solution with AdS asymptotics, and we calculated the null geodesics analytically, in order to determine if there are orbits for the photons that are not allowed in the background of a BTZ black hole. Thus, the differences observed with respect to the BTZ metric are attributed to the breaking of the Lorentz invariance. 

The main difference is that unstable circular orbits are allowed in the Ho\v{r}ava spacetime, which are not allowed in the BTZ spacetime. Consequently, there are two critical orbits that approach to this unstable circular orbit asymptotically. In the critical orbit of first kind, the particle arises from infinity, and in the critical orbit of second kind, the particle starts from a finite distance bigger than the horizon radius, but smaller than the unstable radius. Other orbit that is not allowed in the BTZ spacetime is the orbit of first kind, which implies that deflection of the light is possible in the background analyzed.
 
 We determined the analytical trajectory of the photons corresponding to the bending of the light, and we calculated exactly the angle of deflection. Also, for radial geodesics, we showed that as seen by a system external to photons, they will fall asymptotically to the event horizon. On the other hand, this external observer will see that photons arrive in a finite coordinate time to the spatial infinity. Remarkably, this behavior is similar to the observed one in Lifshitz's spacetimes \cite{Villanueva:2013gra}.

\section*{Acknowledgments}

Y.V. acknowledge support by the Direcci\'on de Investigaci\'on y Desarrollo de la Universidad de La Serena, Grant No. PR18142. P.A.G. would like to thank the Facultad de Ciencias, Universidad de La Serena for its hospitality. Y.V. would like to thank the Facultad de Ingenier\'{i}a y Ciencias, Universidad Diego Portales for its hospitality.  P. A. G. and M.O. acknowledge the hospitality of the National Technical University of Athens, where part of this work was undertaken.

%\appendix

\end{document}